\documentclass[12pt]{iopart}
\usepackage{amssymb}
\usepackage{graphicx}
\usepackage[dvips]{epsfig}
\usepackage{subfigure}
\usepackage{bm}

\input{tcilatex}

\begin{document}

\title[Correlated interactions in a biological 
coevolution model]{ Effects of correlated interactions in a biological 
coevolution model with individual-based dynamics }
\author{Volkan Sevim$^1$ and Per Arne Rikvold$^{1,2}$}
\address{
$^1$ School of Computational Science,
Center for Materials Research and Technology,
and Department of Physics, 
Florida State University, Tallahassee, FL 32306-4120, USA\\
$^2$ National High Magnetic Field Laboratory,
Tallahassee, FL 32310-3706, USA}
\eads{\mailto{sevim@csit.fsu.edu}, \mailto{rikvold@csit.fsu.edu} }

\begin{abstract}
Models of biological coevolution have recently been proposed and
studied, in which a species is defined by a genome in the form
of a finite bitstring, and the interactions between species
$i$ and $j$ are given by a fixed matrix with independent, 
randomly distributed elements $M_{ij}$. A consequence of the stochastic 
independence is that species whose genotypes differ even by a single bit may have completely
different phenotypes, as defined by their interactions with the other species. This is clearly
unrealistic, as closely related species should be
similar in their interactions with the rest of the ecosystem. 
Here we therefore study a model, in which the $M_{ij}$  
are correlated to a controllable degree by means of a local averaging scheme. 
We calculate, both analytically and numerically,
the correlation function for matrix elements $M_{ij}$
and $M_{kl}$ versus the Hamming distance between the bitstrings
representing the species. The agreement between the analytical and numerical calculations is excellent for correlations of limited range, 
but explainable differences arise for correlation ranges that are a significant fraction of the length of the bitstring. 
We compare long kinetic Monte Carlo
simulations of coevolution models with uncorrelated and
correlated interactions, respectively. In particular, we consider
the probability density for the lifetimes of individual species.
The species-lifetime distribution is close to a
power law with an exponent near $-2$ over eight decades 
in time in both uncorrelated and correlated cases. The durations of quasi-steady states and power spectral densities for the diversity indices display noticeable differences. However, some qualitative features, like $1/f$ behaviour in power spectral densities for the diversity indices, are not affected by the correlations in the interaction matrix.
\end{abstract}

\pacs{
87.23.Kg, 
05.40.-a, 
05.65.+b, 
89.75.-k 
} 
\submitto{Physical Biology}

\maketitle
\section{Introduction}
\label{sec:I}

During the last decade, ecological dynamics and biological evolution 
have become new areas of 
interest for statistical physicists~\cite{Drossel:2001}. This trend 
has merged with recent 
developments in the theory of complex networks \cite{ALBE02}, 
which has drawn increased attention to problems in ecological dynamics, 
such as the evolution and stability of food 
webs~\cite{Drossel:2001, Drossel:2002,DUNN02,GARL03,GARL04}. 

While many models of macroevolution are formulated explicitly at
evolutionary (macroscopic) time scales, evolutionary dynamics in nature 
are driven by reproduction, mutations, and selection at the 
ecological (microscopic) scale of individual organisms. Although
this means that individual-based models of macroevolution must span
a dauntingly large range of timescales, this is now becoming
possible with the aid of modern computers and algorithms. 
We therefore recently studied an individual-based  
biological coevolution model~\cite{Rikvold:2003,Zia:2004,RIKV05A} 
with random interspecies interactions.
This model is a simplified version of the tangled-nature model 
introduced by Jensen, 
{\it et al.}~\cite{Hall:2002,Christensen:2002,COLL03}. 
Individuals are represented by a haploid genome 
consisting of a string of bits. 
The individuals reproduce asexually and interact with individuals
of other species through 
a random interaction matrix with stochastically independent
elements, which determine how the reproduction
probability of individuals of 
each species is affected by the presence and abundance of 
other species in the ``ecosystem". During reproduction, an
offspring individual can randomly mutate with a 
small but constant probability by having a single bit in its 
genome flipped. These mutations create perturbations in the
ecosystem by introducing new species. A particular mutant may or may not 
succeed, depending on its interactions with the species
already present in 
the system. Usually a mutant does not fit in well with the resident
species, and so it quickly goes extinct. However, occasionally a
successful mutant 
can cause avalanches of mass extinctions, destroying some or all 
of the other species through predation or competition. 
As a consequence, the individuals in this model ecosystem 
live in a dynamic ``fitness landscape", which evolves due to changes 
in the populations of the resident species and the introduction of 
new species through mutations. 
The model displays quiet periods interrupted by bursts of high activity 
caused by mass extinctions that are triggered by the introduction of new 
mutants, reminiscent of the punctuated-equilibrium mode of evolution 
suggested by Eldredge and Gould~\cite{ELDR72,Gould:1977, Gould:1993}. 
The lifetime distribution for individual species is well described by a 
power law with exponent close to $-2$. 

A rather unrealistic aspect of this model is the fact that the elements of
the interaction matrix are completely uncorrelated. This means that
a mutant in general will have completely different interactions
with other species, and thus a completely different phenotype, than does its parent ``wildtype". 
In the present paper, we therefore modify our previous 
model by introducing correlations into the interaction matrix. 
(Similar ideas have been pursued by 
Kauffman~\cite{Kauffman:1991, originsoforder}.) 
In the modified model, the correlations between matrix elements
ensure that 
mutants are not completely different from their parents. Rather, 
their interactions with the other species are similar to those of the
wildtype. The price to be paid is a significant reduction in the
number of effectively independent phenotypes for a genome of a given length. 
In this paper we compare numerical results for this 
modified model with results for
the corresponding 
uncorrelated model in order to assess the effects of including correlations. 
The results show that, although long-range correlations affect the dynamics of the system noticeably, some qualitative features remain unchanged. 

The organization of the rest of this paper is as follows. 
In section~\ref{sec:M} we describe the model and the algorithm used 
for the simulations. In section~\ref{sec:X} we describe the construction 
of the correlated interaction matrix and briefly discuss some 
complications of this process. In section~\ref{sec:RES} we report our 
results, which are summarized in section~\ref{sec:SUM}. Derivations of 
our analytical results are provided in the Appendices. 

\section{Model}
\label{sec:M}
A species in our \index{Monte Carlo}Monte Carlo (MC) simulations is represented by a bitstring genome of length $L$. This $L$-bit genome supplies a pool of $2^L$ possible species. These organisms are considered haploids. Thus, the terms ``species", ``species index", and ``genotype" are synonymous in this paper. 

Individuals in the simulation are allowed to give birth to $F$ offspring 
per generation with a probability $P_i$, where $i$ denotes the species 
index. The reproduction is asexual (cloning), and it occurs in discrete, 
nonoverlapping generations. Thus, only offspring can survive to the next 
generation, and all individuals die at the end of the generation whether 
they reproduce or not. We use a mathematically convenient, nonlinear form for the probability of 
reproduction 
\cite{Rikvold:2003,Zia:2004,Hall:2002,Christensen:2002,Sevim:2004}, 
given by

\begin{equation}
P_i(\{n_j(t)\})
=
\frac{1}{1 + \exp\left[ - \sum_j M_{ij} n_j(t)/N_{\rm tot}(t)
+ N_{\rm tot}(t)/N_0 \right]}
\;,
\label{eq:P}
\end{equation}
where $n_i(t)$ is the population of species $i$ at time $t$. The Verhulst factor $N_0$ represents the carrying capacity of the ``ecosystem" and keeps the total population, $N_{\rm tot}(t) = \sum_i n_i(t)$, finite. $\bf M$ is the interaction matrix, in which a matrix element $M_{ij}$ represents the effect of the population density of species $j$ on species $i$. Species $i$ benefits from species $j$ if $M_{ij}$ is positive, and it is inhibited by species $j$ if $M_{ij}$ is negative. There is no interaction between species $i$ and $j$ when $M_{ij}$ is zero. The structure of $\bf M$ is discussed in the next section.

Although we begin the simulations with a population consisting of only one species, the system diversifies quickly because, in each generation, all individuals are exposed to mutations that create new species. The mutation process is modelled by flipping a randomly chosen bit in the individual's genome with a probability $\mu$ per generation per individual.

  We note that the population dynamics used in the model is somewhat unrealistic, due to the absence of external energy resources and  conservation of energy (or biomass) \cite{Rikvold:2003, Zia:2004,RIKV05A,Hall:2002,Christensen:2002,COLL03}. 
\begin{figure}[t]
\centering \vspace{0.2 truecm} 
\includegraphics[width=.65\textwidth]{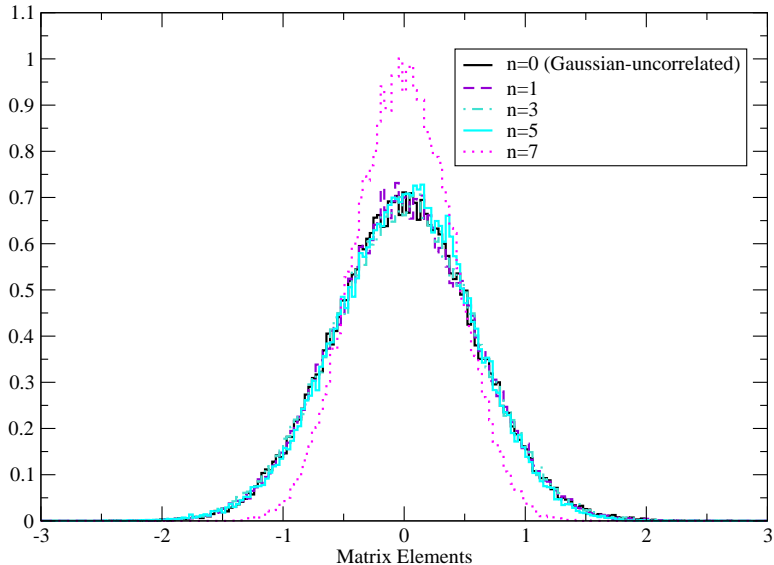}
\caption[]{Some realizations of matrix-element distributions for typical runs with $L=8$ and $n=0$, 1, 3, 5, and 7. (The distributions are {\it not} averaged over independent runs.) For these particular realizations, the distributions for $n=0$, 1, 3, and 5 practically overlap. The distribution for $n=7$ is narrower due to effects that are explained in section~\ref{sec:complications}. }
\label{fig:fig1}
\end{figure}

\section{The Interaction Matrix}
\label{sec:X}
\subsection{Form of the Matrix}
The structure of the interaction matrix characterizes the dynamics of the system. Rikvold and Zia~\cite{Rikvold:2003,Zia:2004} studied the same system (except for a negligible probability of $O(\mu^2)$ for multiple mutations) using an interaction matrix with  off-diagonal elements distributed randomly and uniformly over $[-1,1]$ and with the diagonal elements set to zero. We shall call this the uniform-uncorrelated model. For reasons that will become clear later, we instead use a Gaussian distribution to facilitate some comparisons. We call this model the Gaussian-uncorrelated (or just the uncorrelated) model. 

The main focus of this paper is the effects of correlations in the interaction matrix. (Some preliminary results of this work were discussed in reference~\cite{Sevim:2004}.) The motivation behind the introduction of correlations is the following.  In reality, the interaction between two species $x$ and $z$ should be positively correlated with the interaction between species $y$ and $z$ if $x$ and $y$ are closely related. Therefore,  one would expect a positive correlation between the matrix elements $M_{xz}$ and $M_{yz}$.  

A correlated matrix is generated using the following method. First,
a random matrix with elements drawn from a Gaussian distribution with standard deviation $\sigma_0=\sqrt{1/3}$ and mean zero is
generated. (We chose this standard deviation for backward
compatibility with earlier studies in which the matrix elements
were uniformly distributed on $[-1,1]$.) Then, each matrix element
in the random matrix is averaged over itself and its neighbours in genome space up to a Hamming distance $n$, where $n$ is the averaging radius. Thus, the total number of terms in the average (or the number of points inside the averaging ``hypersphere" with radius $n$ in genome space) is  $Z_{n}=\sum_{m=0}^{n}{\binom{2L}{{m}}}$. The average is 
multiplied by $\sqrt{Z_{n}}$ in order to keep the standard deviation of the correlated
matrix elements approximately the same as for the random ones. This process eliminates any possible effect by a change in the shape of the distribution of interaction strengths. Thus, a correlated matrix element $M_{ij}$ is given by 
\begin{equation}
M_{ij}=\sum_{H(ij;kl)\leq n}M_{kl}^{0}/\sqrt{Z_{n}} \;, \label{eq:AV}
\end{equation}%
where $M_{kl}^{0}$ are the elements of the uncorrelated matrix, which are
Gaussian distributed with standard deviation $\sigma_0$, and $H(ij;kl)$ is the Hamming distance
between two matrix elements $M_{ij}^{0}$ and $M_{kl}^{0}$. We use the city-block metric to define the
Hamming distance between two matrix elements. Therefore, $%
H(ij;kl)=H(i,k)+H(j,l)$, where $H(i,j)$ is the Hamming distance between
bitstrings $i$ and $j$. In fact, $H(ij;kl)$ denotes the Hamming distance between the concatenated bitstrings $ij$ and $kl$. (This concatenation notation is useful for Hamming distance calculations and will be used throughout the paper.) The diagonals are set to zero {\it after} the averaging process is finished. The model in which such a correlated interaction matrix is used is called the correlated model. Theoretical calculations of correlation functions are discussed in~\ref{sec:theocorr}.

There are two modelling issues that need to be discussed here. The first one is the interpretation of the mutation process. As we mentioned above, in the uncorrelated model, matrix elements $M_{xz}^0$ and  $M_{yz}^0$ are not correlated, even if $y$ is a mutant of $x$. (We call $y$ a mutant of $x$ if $H(x,y)=1$.) Therefore, the set of interactions between the mutant $y$ and other species can be completely different than those between the wildtype $x$ and the other species. This lack of relationship between the mutant and the wildtype is not very realistic.
Therefore, the mutation process in the uncorrelated model may, in some sense, be interpreted as the introduction of a completely new species to the system from a pool of species, reminiscent of migration. In the correlated model, the interaction constants of the mutant and the wildtype are correlated. However, whether these correlations are strong enough to describe those between a real mutant and its wildtype is arguable.  The other issue is the non-antisymmetrical form of $\bf M$, meaning $M_{ij} \neq -M_{ji}$. This feature of the model, along with the absence of an external energy source, gives rise to unrealistic numbers of mutualistic interactions in the system~\cite{Rikvold:2003}. 

We have already made a preliminary comparison between the correlated model and the Gaussian- and uniform-uncorrelated models in reference~\cite{Sevim:2004} for short-range correlations, $n=1$ and $n=2$. The results suggested that the correlated and Gaussian-uncorrelated models behave quite similarly, so that short-range correlations in {\bf M} may not have a significant effect on the dynamics. In this paper, we extend our investigation to the effects of longer-range correlations with $n>2$.
\subsection{Distributions of Matrix Elements, Correlations Between Matrix Elements, and Complications in the Averaging Process}
\label{sec:complications}
The averaging process mentioned above causes a few unexpected complications. The first one is a shift in the mean. Although the matrix elements are drawn from a distribution with mean zero, for a finite matrix, the mean, $\langle{M_{ij}^0}\rangle$, while quite small, is always non-zero. The averaging alters the mean of the matrix elements, so that the means of the initial random and the averaged matrices can be considerably different. The reason for the shift is the following. Since each averaged element is multiplied by $\sqrt{Z_{n}}$ to keep the standard deviation constant, the mean of the elements of the averaged interaction matrix becomes $\langle{M_{ij}}\rangle=\sqrt{Z_{n}}\langle{M_{ij}^0}\rangle$. Although $\langle{M_{ij}^0}\rangle$ is a negligibly small number, multiplication by a large $\sqrt{Z_{n}}$ could cause a considerable shift, $\langle{M_{ij}}\rangle - \langle{M_{ij}^0}\rangle$, in the mean. In order to prevent this, we adjust the mean, $\langle{M_{ij}^0}\rangle$, by subtracting  $\langle{M_{ij}^0}\rangle/2^{2L}$ from each element of the initial random matrix. This modification minimizes the shift in the mean.

The other complication of the averaging process is a change in the standard deviation of $M_{ij}$, even though we intend to keep it constant. As in the problem of the shift in the mean mentioned above, the standard deviation of the averaged matrix can be considerably different than that of the initial random matrix. As seen in figure~\ref{fig:fig1}, the matrices averaged for $n=1$ have approximately the same Gaussian form with a standard deviation close to $\sigma_0$, which practically overlaps with the $M_{ij}^0$ distribution of the initial Gaussian-uncorrelated matrix. However, we found that long-range averaging ($n \gtrsim L/2$) could generate distributions with significantly different standard deviations, like the $n=7$ curve included in figure~\ref{fig:fig1}. The reason for the change in the standard deviation is very similar to that for the shift in the mean mentioned above. This anomaly is discussed further in~\ref{sec:APPB}.  

We also checked the correlations between matrix elements to see the effect of averaging. The correlations between two matrix elements,  $M_{ij}$ and $M_{i'j'}$, depend on the overlap of the $M_{kl}^{0}$ terms that occur in the average in equation~(\ref{eq:AV}). Theoretical calculations predict decaying correlation functions with steps due to the city-block metric, as shown in figure~\ref{fig:theocorrs}. These calculations are explained in~\ref{sec:theocorr}.

\begin{figure}[t]
\centering \vspace{0.3truecm} 
\includegraphics[width=.60\textwidth]{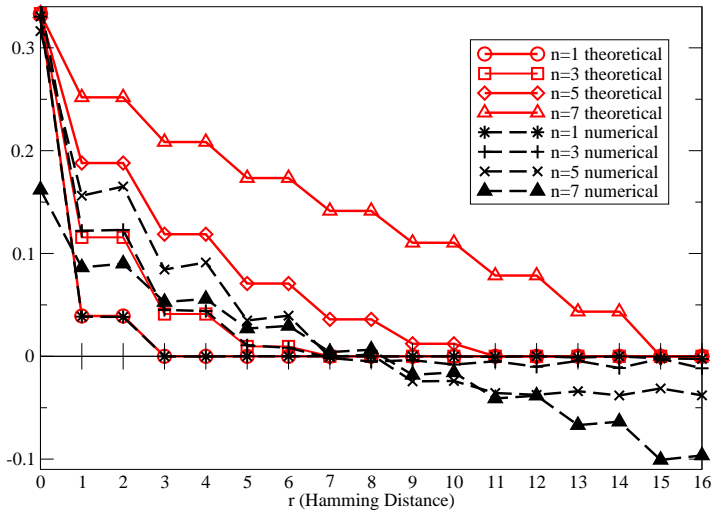}	
\caption[]{Theoretical and some numerical realizations of correlation functions between matrix elements as functions of the Hamming distance between their indices, $r=H(ij,kl)$,   with $L=8$; $n=1$, 3, 5, and 7. The numerical correlation functions were not averaged over independent realizations in order to show the magnitudes and the character of the deviations. Theory and numerics agree for small $n$. Increasing $n$  leads to deviations from the theory, usually retaining the overall shape of the expected correlation function. (Numerical calculations were performed before setting the diagonal of {\bf M} to zero.) }
\label{fig:theoandnumcorrs}
\end{figure}

The correlation functions of the matrices with short-range correlations are in good agreement with the theory. However, as $n$ is increased, the numerical correlation functions begin to deviate from their theoretical counterparts, as shown in figure~\ref{fig:theoandnumcorrs}. The numerical correlation functions of the highly correlated matrices are distorted and translated along the $y$-axis, compared to the theoretical ones. We believe that the cause of this anomaly is related to the cause of the anomaly in the standard deviations. This complication is discussed in~\ref{sec:APPB}.

\begin{figure}[t]
\centering \vspace{1.3truecm} 
\includegraphics[width=.85\textwidth]{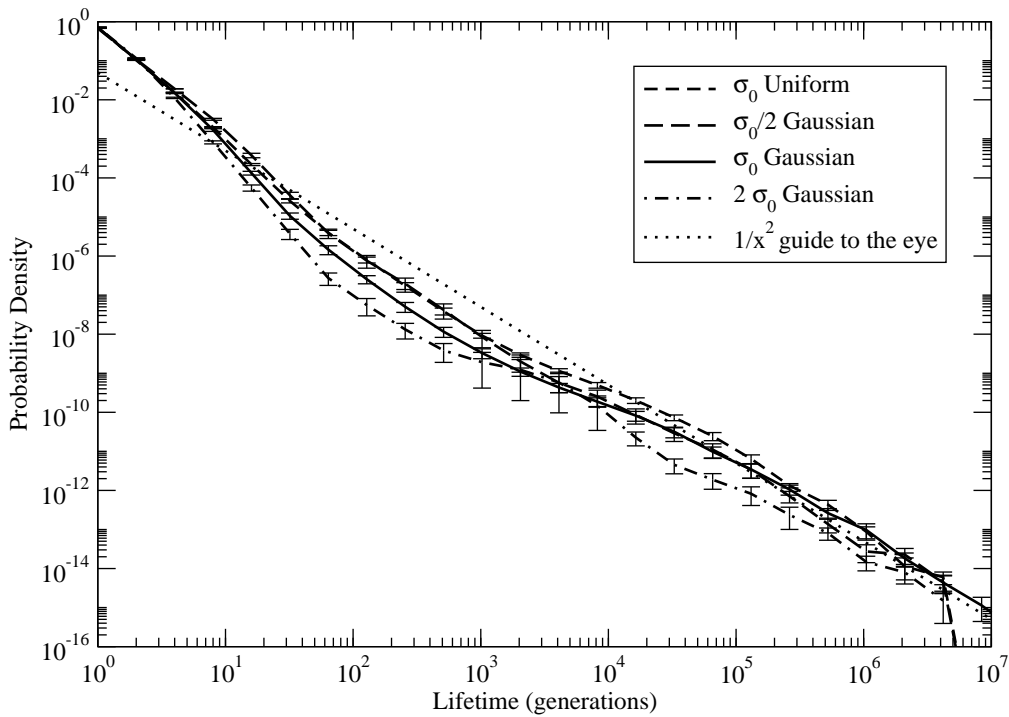}	
\caption[]{Normalized histograms of species-lifetimes for the uncorrelated 
model with a uniform matrix-element distribution with standard deviation $\sigma_0$, and with  Gaussian matrix-element distribution widths, $\sigma_0/2$, $\sigma_0$,  
and $2\sigma_0$. Based on simulations of $2^{25}$ generations each with $L=8$, averaged over eight runs.}
\label{fig:lifetime-uncorr}
\end{figure}

\begin{figure}[t]
\centering \vspace{0.3truecm}
\includegraphics[width=.90\textwidth]{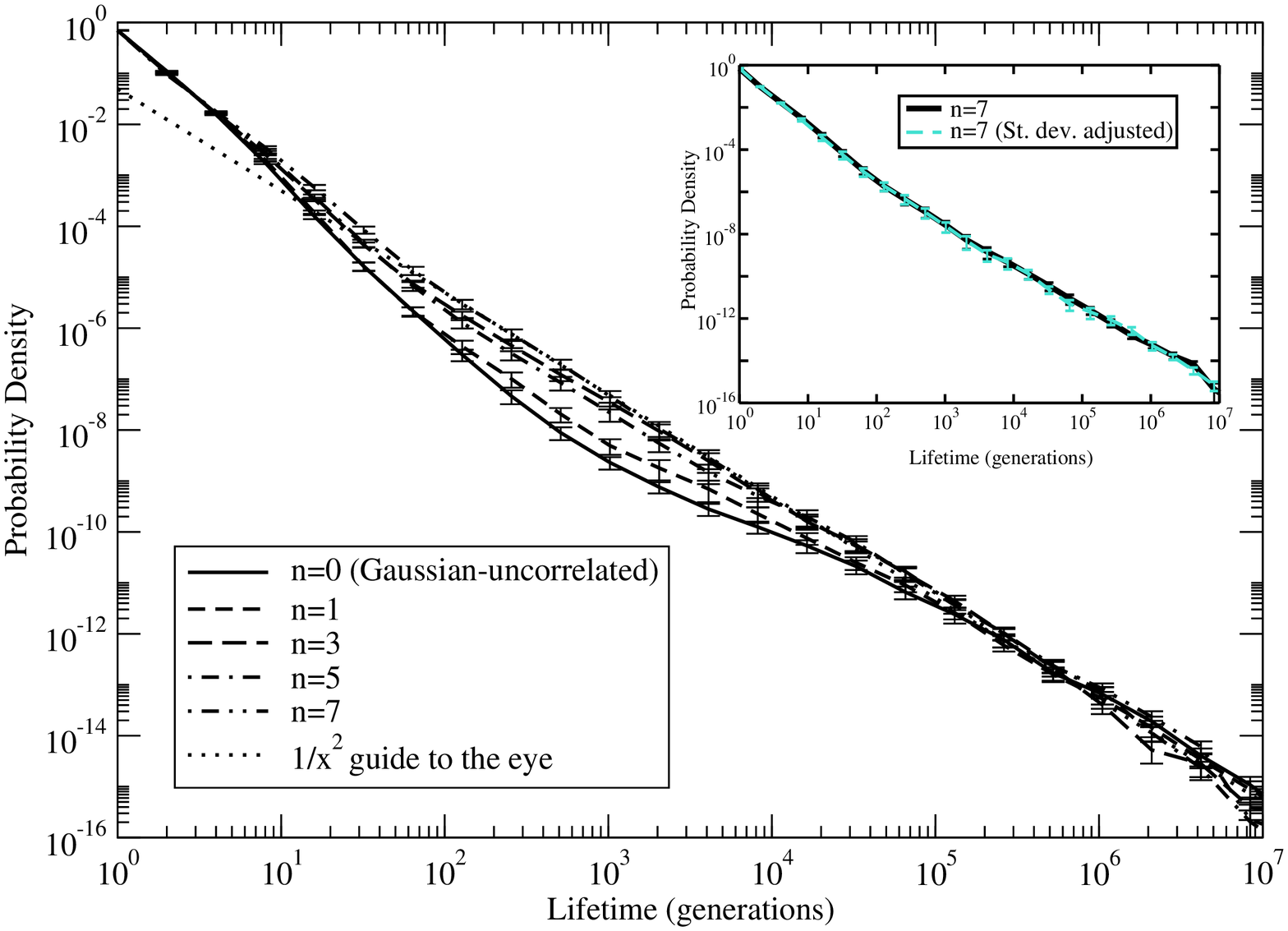}	
\caption[]{Normalized histograms of species lifetimes for uncorrelated and correlated models with different averaging radii, based on 
simulations of $2^{25}$ generations each with $L=8$. Each curve represents an average over sixteen runs. The standard deviations of the $M_{ij}$ distributions change after averaging.
The averages of the new standard deviations are $0.5755(6)$, $0.570(3)$, $0.54(1)$, and $0.46(1)$ for $n=1$, 3, 5, and 7, respectively, compared to $\sigma_0=0.5773$. Inset: The lifetime histograms for $n=7$ with and without adjusting the standard deviation of the $M_{ij}$, averaged over eight runs each. The same sets of random numbers were used for the ``adjusted" eight-run set as for their ``non-adjusted" counterparts. The distributions practically overlap, suggesting that the change in the standard deviation has much less effect on the lifetime distribution than the correlations in {\bf M}.}
\label{fig:lifetime}
\end{figure}

\section {Simulation Results}
\label{sec:RES}
Unless otherwise indicated, we performed sixteen independent runs  for $2^{25}$=33\,554\,432  generations for each model using parameters similar to those in references~\cite{Rikvold:2003} and~\cite{Sevim:2004}: genome length $L=8$ bits,
mutation rate $\mu = 10^{-3}$ per individual per generation,
carrying capacity $N_0 = 2000$, fecundity $F=4$; and for the correlated model, averaging radii $n=1, 3, 5,$ and 7. Each run used a different $\bf M$ and a different sequence of random numbers. Each simulation started with 200 individuals of genotype 0. The system parameters were recorded every sixteen steps. 

In order to see the effect of the shape of the $M_{ij}^0$ distribution, we first ran four sets of simulations of the uncorrelated model with a uniform $M_{ij}^0$ distribution with standard deviation  $\sigma_0$, and with a Gaussian $M_{ij}^0$ distribution with standard deviations $\sigma_0/2$, $\sigma_0$ and $2\sigma_0$. For these systems with different $M_{ij}^0$ distributions, we compared histograms of species lifetimes. The species lifetime is defined as the number of generations between creation and extinction of a species, during which its population is continuously positive. 

As seen in figure~\ref{fig:lifetime-uncorr}, the lifetime distributions of the uncorrelated systems with different matrix-element distributions all exhibit approximately power-law like decay over seven decades in time, with an exponent near $-2$. Figure~\ref{fig:lifetime-uncorr} also shows that a narrower $M_{ij}^0$ distribution flattens the concavity between $10$ and $10^4$ generations in the lifetime distribution, while a wider one makes it deeper.  

The lifetime distributions of the correlated models shown in figure~\ref{fig:lifetime} also exhibit a power-law like decay similar to the ones in figure~\ref{fig:lifetime-uncorr}. The distributions obtained from the simulations with very short-range correlations, like $n=1$, practically overlap with the one corresponding to the uncorrelated model. Increasing $n$ leads to a deviation from the distribution of the uncorrelated model. The $n=3, 5$, and 7 curves in figure~\ref{fig:lifetime} look similar to the $\sigma_0/2$ curve in figure~\ref{fig:lifetime-uncorr}, suggesting that the change in the standard deviation (or the shape) of the $M_{ij}$ distribution could perhaps be responsible for this anomaly. In order to make sure that it is actually the correlations that cause the changes, we suppressed this anomaly: we ran another set of simulations for $n=7$ in which the standard deviation of the $M_{ij}$ distribution was adjusted to $\sigma_0$ after averaging. The result is shown in the inset in figure~\ref{fig:lifetime}. 
The two lifetime distributions in the inset almost completely coincide. They are both the averages of eight-run sets with $n=7$ (using the same random number sequences), except that the $M_{ij}$ distribution of the second set is adjusted after averaging. The very small difference between these two curves shows that the effect of a change in the standard deviation of the $M_{ij}$ distribution after averaging is negligible. Thus, the correlations in $\bf M$ are largely responsible for the differences between the lifetime distributions of the correlated and uncorrelated models, shown in the main part of figure~\ref{fig:lifetime}.

The deviation trend in the lifetime distributions is not monotonic in the averaging radius (the $n=5$ curve appears below  the $n=3$ curve). However, the error bars for the $n=3$ and $n=5$ curves overlap. The large error margin due to the anomaly in the correlation functions leaves some uncertainty. Nevertheless, even fairly long-range correlations in $\bf M$ (like $n=5$ and 7) do not change the gross features like the approximate $1/t^2$ behaviour of the lifetime distributions.

\begin{figure}[t]
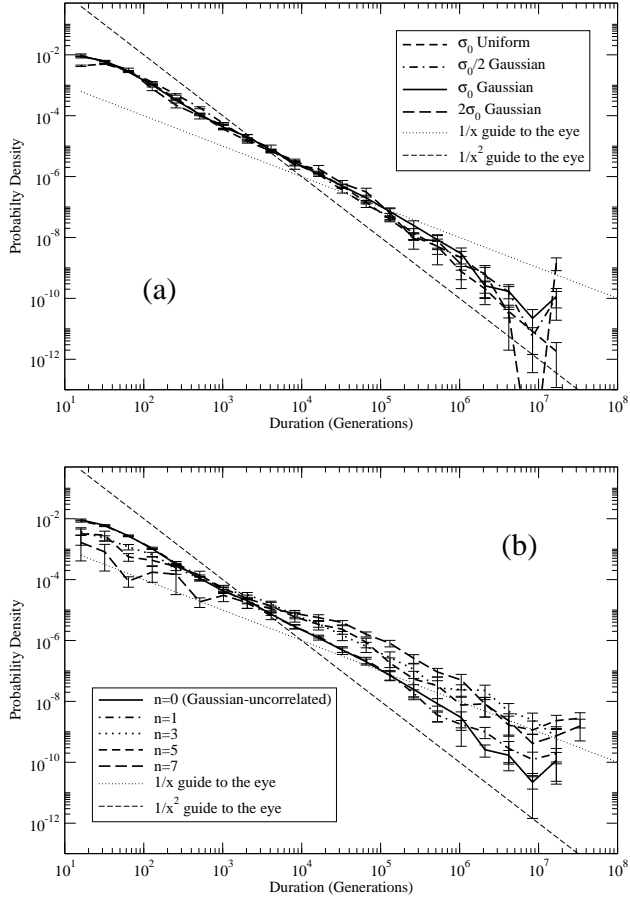

\centering
\includegraphics[width=.57\textwidth,viewport= 0mm 0mm 255mm 175mm ,clip]{qss-uncorr.eps}
\vspace{0.6truecm}
\includegraphics[width=.57\textwidth,viewport= 0mm 0mm 255mm 175mm ,clip]{qss-corr.eps}		
\caption[]{Normalized histograms for the duration of quasi-steady states. A cut-off of $|S(t) - S(t-16)|/16=0.015$ was used to distinguish between QSS and active periods. Each curve represents an average over eight runs for uncorrelated models and sixteen runs for correlated models. Based on simulations of $2^{25}$ generations each, sampled every sixteen generations. 
(a) Simulations for uncorrelated models with uniform and Gaussian $M_{ij}$ distributions with the standard deviations $\sigma_0$, and $\sigma_0/2$, $\sigma_0$, and $2\sigma_0$, respectively. The distributions are very similar for the uncorrelated model. (b) Simulations for the uncorrelated and the correlated models with $n=1, 3, 5,$ and $7$. The Gaussian $\sigma_0$ and the $n=0$ curves are identical. The distributions for the correlated systems differ from the uncorrelated one.}
\label{fig:qss}
\end{figure}

Another set of statistical measurements that we made for the durations of the quasi-steady states (QSS) gave a somewhat conflicting result. The QSSs are the metastable periods, during which the community structure appears to be stationary~\cite{Rikvold:2003,Zia:2004}. They are interrupted by active periods, during which the community structure changes after the emergence of a successful mutant and resulting extinction of other species. In order to identify the QSSs, we recorded the entropy of the system every sixteen generations. The information-theoretical entropy is given by \cite{SHAN48,SHAN49},
\begin{equation}
S\left( \{ n_I(t) \} \right) 
=
- \sum_{\{I | \rho_I(t) > 0 \}} \rho_I(t) \ln \rho_I(t)
\;,
\label{eq:S}
\end{equation}
where $\rho_I(t) = n_I(t) / N_{\rm tot}(t)$. 
The system is considered in a QSS, as long as the magnitude of the entropy difference, $|S(t) - S(t-16)|/16$, is below a certain threshold, which we took as $0.015$. (For an extensive discussion on how to obtain this value, see reference~\cite{Rikvold:2003}.) The QSS-duration distribution for the uncorrelated model is not significantly affected by the change in the standard deviation of the $M_{ij}$ distribution, except the density of the longest lived communities around $10^7$ generations (figure~\ref{fig:qss}(a)). Considering the large error bars at this range, the QSS distributions of the uncorrelated systems nearly overlap. However, the correlations in $\bf M$ seem to increase the lifetimes of the QSS communities (figure~\ref{fig:qss}(b)). 

\begin{figure}[t]
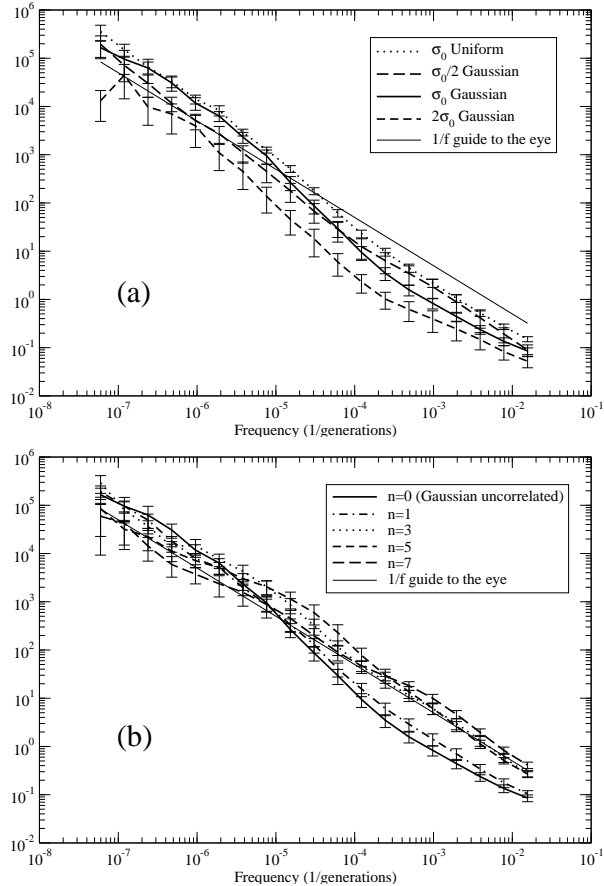

\centering
\includegraphics[width=.55\textwidth,viewport= 1mm 1mm 255mm 175mm ,clip]{psds-uncorr.eps}	
\vspace{0.6truecm}
\includegraphics[width=.55\textwidth,viewport= 1mm 1mm 255mm 175mm ,clip]{psds-corr.eps}	
\caption[]{Power spectral densities of Shannon-Wiener diversity indices for different models. Each curve represents an average over eight runs for the uncorrelated models and sixteen runs for the correlated models. Based on simulations of $2^{25}$ generations each,  sampled every sixteen generations. The results obtained from each run were also averaged over each octave to reduce the noise. 
(a) Simulations for uncorrelated models with uniform and Gaussian $M_{ij}$ distributions with the standard deviations $\sigma_0$, and $\sigma_0/2$, $\sigma_0$, and $2\sigma_0$, respectively.
(b) Simulations for uncorrelated and correlated models with $n=1, 3, 5,$ and $7$.  The Gaussian $\sigma_0$ and the $n=0$ curves are identical. The distributions for the correlated systems differ slightly from the uncorrelated one.}
\label{fig:psd}
\end{figure}

In order to get more information on how the fluctuations in the system are affected by the correlations in $\bf M$, we also calculated the power spectral density (PSD) of the Shannon-Wiener diversity index. The Shannon-Wiener diversity index~\cite{KREB89} is defined as $D(t) = \exp \left[S \left( \{ n_I(t) \} \right) \right]$, 
where $S$ is the information-theoretical entropy defined in equation~(\ref{eq:S}). In contrast to the QSS statistics (figure~\ref{fig:qss}(a)) (but consistent with the case for the species-lifetime, figure~\ref{fig:lifetime}), the PSDs of the Shannon-Wiener indices of the uncorrelated systems with different $M_{ij}$ distributions do not overlap (figure~\ref{fig:psd}(a))~\cite{PSD,PRES92}. Although the PSDs are not monotonic in the standard deviation of $M_{ij}$, they have a similar $1/f$ like shape, indicating that the general characteristics of the system do not vary significantly with the $M_{ij}$ distribution. However, the correlations in $\bf M$ lead to a more pronounced $1/f$ like behaviour in the PSDs by increasing the relative weight of high-frequency fluctuations as the correlations grow stronger (figure~\ref{fig:psd}(b)). Although the deviations from the uncorrelated case are neither gradual nor monotonic in correlation strength, the result suggests that the correlations in $\bf M$ may in fact change some characteristics of the system, supporting the result obtained from the QSS duration statistics. Nevertheless, the qualitative $1/f$ behaviour of the PSDs is not affected by these changes.

The results discussed above may seem rather complicated, but we
nevertheless believe some clear effects can be discerned. 

There are clear 
differences between the results for the correlated and
uncorrelated Gaussian matrix elements. The correlations further
increase the proportion of very long QSS (figure~5b), but
curiously they also lead to an increase of intermediate-lifetime
species (figure~4). These effects do not seem very easy to
reconcile, but we speculate that they may be due to a decoupling of
the species lifetimes and the QSS durations that has been observed
in an uncorrelated
predator-prey version of this model [9]. The increased activity in
the intermediate-time regime results in increased intensity in the
higher-frequency half of the PSD (figure~6b). 
It should be noted that strong correlation effects are only
seen when $n$ becomes a significant proportion of the genome length
$L$ (here, typically $n \ge 3$). Such long-range correlations are of
very limited interest for the biological systems,
and they only become significant in the present study
because we, for computational reasons, are working with the
unrealistically short genome length, $L=8$. With $n \ll L$, the
effects of correlations appear to be quite minor. 


\section {Summary and Conclusions}
\label{sec:SUM}

In this paper we have considered the effects of introducing
correlations between the elements of 
the interaction matrix $\bf M$ that determines the
interspecies interactions in an individual-based coevolution model 
\cite{Rikvold:2003,Zia:2004,RIKV05A}, in which individuals are
represented by a genome in the form of a bitstring of length $L$. 

The correlations were introduced by replacing each element
$M^0_{ij}$ in an
uncorrelated interaction matrix with the average over all the
elements $M^0_{kl}$ such that the Hamming distance between the
concatenated bitstrings $ij$ and $kl$ is less than or equal to
$n$, and then reweighting the resulting matrix element $M_{ij}$ such
as to maintain the standard deviation of its probability density
unchanged. 

In section~\ref{sec:X} we calculated numerically for different $n$
the correlation
functions between modified matrix elements $M_{ij}$ and $M_{kl}$ as
functions of the Hamming distance between $ij$ and $kl$, and we
compared these with the theoretical results derived in
\ref{sec:theocorr}. For short-range correlations 
($n < L/2$) we found good
agreement between the numerical and theoretical results, but for
larger $n$ there were significant discrepancies. These
discrepancies were explained in \ref{sec:APPB} 
as a result of the reweighting of the correlated matrix elements
that was performed in order to keep their standard deviation
approximately equal to that of the uncorrelated ones.

Next, in section~\ref{sec:RES}, we performed long kinetic Monte Carlo
simulations of our coevolution model, both with correlated and
uncorrelated interaction matrices. As an indicator of the
similarity of the evolution processes we calculated the probability
densities of the lifetimes of individual species in the two models.
While we found statistically significant differences between the
lifetime distributions in the two models 
(see figure~\ref{fig:lifetime}), in
both cases the distributions stayed near a $1/t^2$ power law over near
seven decades in time. However, the distributions for the durations of  QSSs (figure~\ref{fig:qss}) and the PSDs for the diversity index (figure~\ref{fig:psd}) showed that the uncorrelated and correlated systems have some different characteristics. Nevertheless, the overall behaviour of the system does not seem to change drastically, since the $1/f$ behaviour in the PSDs remain the same. This indicates that the correlations in $\bf M$  affect the long-term behaviour of the system in minor, but rather complicated ways. The correlation effects are sufficiently mild that it is permissible to draw conclusions from simulations of uncorrelated models.


\section*{Acknowledgments}
We thank G. Brown, D. T. Robb, S. Frank, and R. K. P. Zia for helpful comments. This research was supported by U.S.\ National Science Foundation Grant Nos.
DMR-0240078 and DMR-0444051, and by Florida State University through the School of
Computational Science, the Center for Materials Research and Technology, and the National High Magnetic Field Laboratory.

\appendix
\section{Calculating the Correlation Function}
\label{sec:theocorr}  
The correlations between elements of the interaction matrix are
described by the correlation function $C(r)=\langle {M_{ij}M_{kl}}\rangle _{r}-\langle {M_{ij}}\rangle
\langle {M_{kl}}\rangle $. The subscript $r$ means that the averages are not
calculated over all matrix elements, but only over ones that are separated by a Hamming
distance $H(ij;kl)=r$. Assuming the average of the matrix elements, $\langle {M_{ij}}%
\rangle$, is zero, the correlation function simply becomes 
\begin{equation}
C(r)=\langle {M_{ij}M_{kl}}\rangle _{r}\;.  \label{eq:CORR}
\end{equation}%
Substituting equation~(\ref{eq:AV}) into equation~(\ref{eq:CORR}) gives 
\begin{equation}
{\langle {M_{ij}M_{kl}}\rangle _{r}=Z_{n}^{-1}\left\langle \sum_{\rho \leq
n}M_{ab}^{0}\sum_{\rho ^{\prime }\leq n}M_{cd}^{0}\right\rangle _{r}} 
\;,
\label{eq:CORR2}
\end{equation}%
where $\rho =H(ij;ab)$ and $\rho ^{\prime }=H(kl;cd)$. By expanding the sums
in the average, we obtain 
\begin{equation}
\langle {M_{ij}M_{kl}}\rangle _{r}=\left\langle (M_{ab}^{0}+M_{a^{\prime
}b^{\prime }}^{0}+\dots )(M_{cd}^{0}+M_{c^{\prime }d^{\prime }}^{0}+\dots
)\right\rangle /Z_{n}
\;.  \label{eq:CORR3}
\end{equation}%
In this average, all cross terms, $\langle {M_{ab}^{0}M_{cd}^{0}}\rangle $
where $a\neq c$ and/or $b\neq d$, are equal to zero. However, if $a=c$ and $%
b=d$, then the term $\langle {(M_{ab}^{0})^{2}}\rangle $ is equal to $\sigma_0^2$,
which is the variance of the elements of the random interaction matrix.
Therefore, the correlation function becomes
\begin{equation}
C(r)=\langle {M_{ij}M_{kl}}\rangle _{r}={\mathcal{N}_{L,n}(r)\sigma_{0}^2}/{Z_n}={\mathcal{N}_{L,n}(r)}/{3Z_n}
\;,
\label{eq:CORR4}
\end{equation}%
where $\mathcal{N}_{L,n}(r)$ is the
number of common terms in the two sums in equation~(\ref{eq:CORR2}). The calculation of $\mathcal{N}_{L,n}(r)$ is explained below. 
\begin{figure}[t]
\centering \vspace{0.3truecm} 
\includegraphics[width=.60\textwidth]{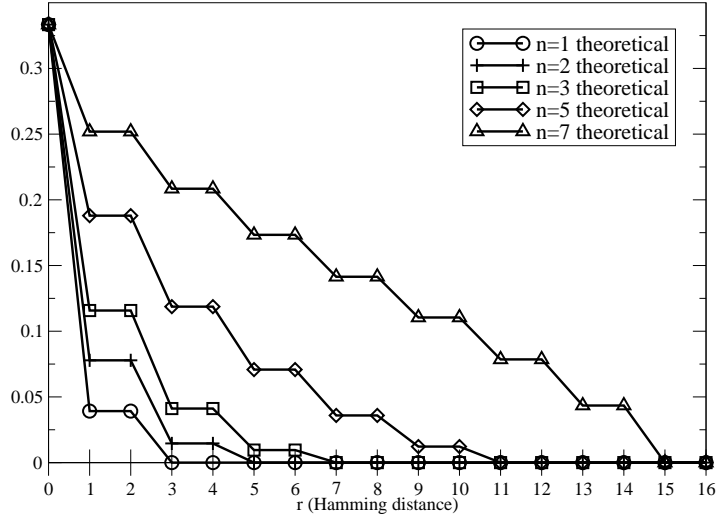}	
\caption[]{Theoretical correlation functions for $L=13$ with $n=1$, 2, 3, 5, and 7.}
\hspace{0.35truecm}
\label{fig:theocorrs}
\end{figure}


\begin{figure}[t]
\centering \vspace{0.2truecm} 
\includegraphics[angle=270, width=.80\textwidth, viewport=12mm 20mm 80mm 255mm,clip]{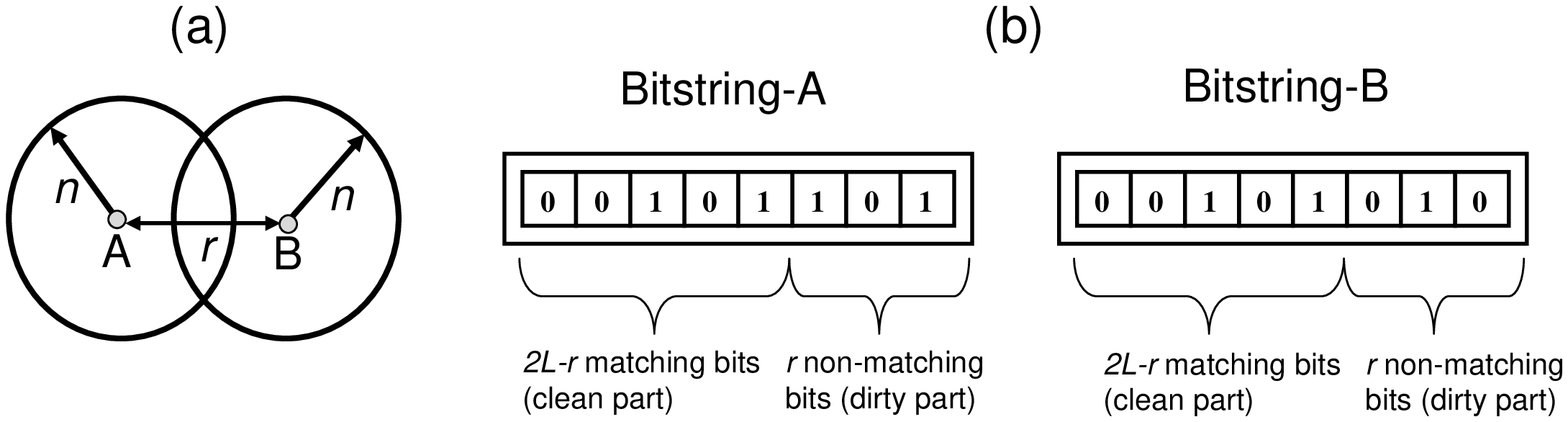}
\caption[]{(a) Common elements in the two sums shown symbolically as the
intersection of two $2L$-dimensional hyperspheres in genome space with
radii $n$. (b) ``Dirty" and ``clean" parts. See discussion in the text.}
\label{fig:dirty-clean}
\end{figure}
\subsection{Calculating $\mathcal{N}_{L,n}(r)$}

\label{sec:CALCN} The terms in the sum in equation~(\ref{eq:AV}) are
neighbours of $M_{ij}^{0}$ within a hypersphere of ``radius" $n$, centred at $ij$ in
the $2L$-dimensional genome space. So, the common terms in the sums
in equation~(\ref{eq:CORR2}) are the ones which lie in the intersection of two $2L$-dimensional ``hyperspheres" of radii $n$, centred at $A=ij$ and $B=kl$ (figure~\ref{fig:dirty-clean}(a)). This analogy helps us to reformulate the problem of
finding the common terms in the sums in equation~(\ref{eq:CORR2}) as the following.
What is the total number of bitstrings that are $n$ bits or less away from
both bitstrings $A$ and $B$, when $A$ and $B$ differ from each other by $r$ bits? Equivalently, we want
to find the number of identical bitstrings we can create by making $n$ or less modifications on $A$ and $B$ each.

To make it easier to visualize the solution, we arrange the two bitstrings $A$ and $B$ so that the matching bits constitute the left-hand part of each
bitstring. We shall call these $2L-r$ matching bits the \textit{clean part}, and the $r$ non-matching bits the \textit{dirty part} (figure~\ref{fig:dirty-clean}(b)). We emphasize that this arrangement is only for visualization purposes, and not a part of the solution.

The contributions to $\mathcal{N}_{L,n}(r)$ can be grouped in three cases:

\begin{itemize}
\item[a)] $n\geq r\geq 0$: First, we make the dirty parts identical by making $i$
modifications on either $A$ or $B$ and $r-i$ modifications on the other.
There are $\sum_{i=0}^{r}{\binom{r}{{i}}}=\sum_{i=0}^{r}{\binom{r}{{r-i}}}%
=2^{r}$ ways of making the dirty parts identical. Since $n>r$, we can also
make $k=n-\max (r-i,i)=\min (n-r+i,n-i)$ modifications on the clean part,
unless $k>2L-r,$ in which case the number of modifications exceeds $2L-r$, the size of the clean part. Therefore, the correct upper limit for $k$ should be 
$\min (2L-r,n-r+i,n-i)$. Thus, the total number of different configurations
that can be obtained on the clean part is ${\sum_{k=0}^{\min
(2L-r,n-r+i,n-i)}{{\binom{2L-r}{{k}}}}}$. Combining these, we obtain the
contribution to $\mathcal{N}_{L,n}(r)$ for the $0\leq r\leq n$ case: 
\begin{equation}
\sum_{i=0}^{r}{\sum_{k=0}^{\min (2L-r,n-r+i,n-i)}{{\binom{2L-r}{{k}}}{\binom{%
r}{{i}}.}}}  \label{eq:Nr1}
\end{equation}

\item[b)] $2n\geq r>n$: When $r>n$, we have to make at least $r-n$
modifications on one of the dirty parts. Therefore, the contribution to $%
\mathcal{N}_{L,n}(r)$ for the $2n\geq r>n$ case is 
\begin{equation}
\sum_{i=r-n}^{n}{\sum_{k=0}^{\min (2L-r,n-r+i,n-i)}{{\binom{2L-r}{{k}}}{%
\binom{r}{{i}}}}} \;. \label{eq:Nr2}
\end{equation}%

Note that the upper limit of the first sum is not $r$ anymore, since we are
allowed to make only $n<r$ modifications in total.
\item[c)] $r>2n$: Since the two hyperspheres in figure~\ref{fig:dirty-clean}(a) cannot overlap when $%
r>2n$, the contribution to $\mathcal{N}_{L,n}(r)$ in this case is zero.
\end{itemize}

We obtain the final form of $\mathcal{N}_{L,n}(r)$ by combining equation~(\ref{eq:Nr1}%
) and equation~(\ref{eq:Nr2}): 
\begin{equation}
\fl
\mathcal{N}_{L,n}(r)=\left\{ 
\begin{array}{ll}
\dsum\limits_{i=\max (r-n,0)}^{\min (n,r)}{\dsum\limits_{k=0}^{\min
(2L-r,n-r+i,n-i)}{{\dbinom{2L-r}{{k}}}{\dbinom{r}{{i}}}}} & 
\mbox{if $r \leq
2n$} \\ 
0 & \mbox{otherwise}%
\end{array}%
\right. .  \label{eq:FINALNr}
\end{equation}

A plot of the theoretical correlation functions for $L=13$ with $n=$1, 2, 3, 5, and 7 is shown in figure~\ref{fig:theocorrs}.


\section{Deviations from the Theory in Correlation Functions of Highly Correlated Matrices
}
\label{sec:APPB} The theoretical calculations for the correlation function agree
with the numerical results for matrices with short-range correlations. However,
the numerically obtained correlation functions for highly correlated matrices
differ significantly from the theory, as seen in figure~\ref{fig:theoandnumcorrs}. We here demonstrate
why these deviations occur by calculating the variance of the correlated
matrix elements (which is equal to the value of the correlation function at $r=0$): 
\begin{equation}
C(0)=\sigma _{\rm corr}^{2}=\langle {(M_{ij})^{2}}\rangle -\langle {M_{ij}}%
\rangle ^{2}.
\end{equation}%
Using 
\begin{equation}
\langle {M_{ij}}\rangle ^{2}=Z_{n}\langle {M_{ij}^{0}}\rangle ^{2}
\end{equation}%
(this time we do not assume $\langle {M_{ij}^{0}}\rangle =0$) and 
\begin{equation}
\fl
\langle {(M_{ij})^{2}}\rangle =\frac{1}{SZ_{n}}\left\{%
(M_{00}^{0})^{2}+M_{00}^{0}M_{01}^{0}+\ldots
+(M_{01}^{0})^{2}+M_{01}^{0}M_{00}^{0}+M_{01}^{0}M_{02}^{0}+\ldots\right\}
\;, 
\end{equation}%
where $S=2^{2L}$ is the total number of matrix elements, we get

\begin{equation}
\fl
\sigma _{\rm corr}^{2}=   
\frac{1}{SZ_{n}}\left\{ Z_{n}\left( (M_{00}^{0})^{2}+(M_{01}^{0})^{2}+\ldots \right)+2\left( M_{00}^{0}M_{01}^{0}+\ldots \right) \right\} -Z_{n}\langle {M_{ij}^{0}}\rangle ^{2} 
\;.
\end{equation}%
Then we transform the terms in parentheses into averages and obtain 
\begin{equation}
\fl
\sigma _{\rm corr}^{2}=\frac{1}{SZ_{n}}\left\{ SZ_{n}\langle {(M_{ij}^{0})^{2}}\rangle +%
\frac{2SZ_{n}(Z_{n}-1)}{2}\langle {M_{ij}^{0}M_{kl}^{0}}\rangle
_{1\leq H(ij;kl)\leq 2n}\right\} -Z_{n}\langle {M_{ij}^{0}}\rangle ^{2}
\;,
\end{equation}%
which simplifies to 
\begin{equation}
\sigma _{\rm corr}^{2}=\langle {(M_{ij}^{0})^{2}}\rangle +(Z_{n}-1)\langle {%
M_{ij}^{0}M_{kl}^{0}}\rangle _{1\leq H(ij;kl)\leq 2n}-Z_{n}\langle {M_{ij}^{0}}\rangle
^{2}\;.  \label{eq:SIGMACORR}
\end{equation}%
Although we intended to keep the variance of the correlated matrix elements
constant through multiplication by $Z_n$ (see equation~(\ref{eq:AV})), equation~(\ref{eq:SIGMACORR}) indicates that the deviations
from the intended theoretical value $\sigma _{0}^{2}=\langle {%
(M_{ij}^{0})^{2}}\rangle $ scale by $Z_{n}$. Our numerical studies showed that
for $Z_{n}\gtrsim L/2$, the term $(Z_{n}-1)\langle {M_{ij}^{0}M_{kl}^{0}}\rangle
_{1 \leq H(ij;kl)\leq 2n}-Z_{n}\langle {M_{ij}^{0}}\rangle ^{2}$ can no longer be
neglected, and deviations from the theoretical value become
significant.

We verified equation~(\ref{eq:SIGMACORR}) by calculating, $\sigma _{\rm corr}^{2}$, $\langle {%
(M_{ij}^{0})^{2}}\rangle$, $\langle {M_{ij}^{0}M_{kl}^{0}}\rangle
_{1 \leq H(ij;kl)\leq 2n}$, and $\langle {M_{ij}^{0}}\rangle ^{2}$ in computer simulations for $L=7$ with $n=3$ and 4, over 100 runs each. Plugging the values for $\langle {%
(M_{ij}^{0})^{2}}\rangle$, $\langle {M_{ij}^{0}M_{kl}^{0}}\rangle
_{1 \leq H(ij;kl)\leq 2n}$ and $\langle {M_{ij}^{0}}\rangle ^{2}$ obtained from the simulation into the RHS of equation~(\ref{eq:SIGMACORR}) gives the same value for $\sigma _{\rm corr}^{2}$ as in the simulation, with an error of $O(10^{-6})$ . This result supports our theory.

Although this calculation shows the deviation only for 
$C(0)=\sigma _{\rm corr}^{2}$, deviations for $r>0$ should be similar due to the relatively smooth change in the overlap of the averaging hyperspheres. Since the overlaps of the averaging hyperspheres are similar for adjacent $r$'s, the deviations also should be similar for the adjacent $C(r)$'s. For example, the deviations at $C(0)$ and $C(1)$ should be similar in magnitude and direction. This argument explains why the numerical correlation functions look like their theoretical counterparts translated along the $y$ axis, rather than randomly scattered around the expected data points.   

\bibliographystyle{iopart-num}
\bibliography{/home/u2/users/sevim/prj/bib/fwrefs,/home/u2/users/sevim/prj/bib/evol} 

\begin{thebibliography}{10}
\bibitem{PSD}
  The PSD of a variable $y(t)$, where $t$ takes $N$ discrete values with an
  increment $\Delta$, is here defined as $Y(f) = (\Delta / N) | \sum_t [y(t) -
  \bar{y}] \exp (2 \pi i f t) |^2$, where $\bar{y}$ is the average of $y(t)$
  and $f = n/(N \Delta)$ with $n = -N/2, ..., N/2$. It was calculated with
  subroutine {\tt spctrm} from Ref.~\protect\cite{PRES92}, using a Welch window
  and variance reduction as described there.
\end{thebibliography}
\end{document}